\documentclass[twocolumn,showpacs,preprintnumbers,amsmath,amssymb]{revtex4}

\usepackage{graphicx}
\usepackage{dcolumn}
\usepackage{bm}

\begin{document}

\preprint{APS/123-QED}

\title{Field-driven successive phase transitions in quasi-two-dimensional frustrated antiferromagnet Ba$_2$CoTeO$_6$ and highly degenerate classical ground states}

\author{Purintorn Chanlert$^1$}
\author{Nobuyuki Kurita$^1$} 
\author{Hidekazu Tanaka$^1$}
\email{tanaka@lee.phys.titech.ac.jp}
\author{Daiki Goto$^2$}
\author{Akira Matsuo$^2$}
\author{Koichi Kindo$^2$}

\affiliation{
$^1$Department of Physics, Tokyo Institute of Technology, Meguro-ku, Tokyo 152-8551, Japan\\
$^2$Institute for Solid State Physics, The University of Tokyo, Kashiwa, Chiba 277-8581, Japan
}

\date{\today}

\begin{abstract}
We report the results of magnetization and specific heat measurements of Ba$_2$CoTeO$_6$ composed of two subsystems A and B, which are magnetically described as an $S\,{=}\,1/2$ triangular-lattice Heisenberg-like antiferromagnet and a $J_1-J_2$ honeycomb-lattice Ising-like antiferromagnet, respectively. These two subsystems were found to be approximately decoupled. Ba$_2$CoTeO$_6$ undergoes magnetic phase transitions at $T_{\rm N1}\,{=}\,12.0$ K and $T_{\rm N2}\,{=}\,3.0$ K, which can be interpreted as the orderings of subsystems B and A, respectively. Subsystem A exhibits a magnetization plateau at one-third of the saturation magnetization for the magnetic field $H$ perpendicular to the $c$ axis owing to the quantum order-by-disorder, whereas for $H\,{\parallel}\,c$, subsystem B shows three-step metamagnetic transitions with magnetization plateaus at zero, one-third and one-half of the saturation magnetization. The analysis of the magnetization process for subsystem B shows that the classical ground states at these plateaus are infinitely degenerate within the Ising model.
\end{abstract}

\maketitle


\section{Introduction}
Frustrated quantum magnets often provide a stage to embody the remarkable macroscopic quantum many-body effect in a magnetic field~\cite{Miyahara,Chubukov,Nikuni}. In general, the frustrated magnets have a highly degenerate classical ground state in a magnetic field. For a Heisenberg triangular-lattice antiferromagnet (TLAF), which is a typical geometrically frustrated magnet, the classical ground state in the magnetic field is infinitely degenerate. 
This is because the number of equations that determine the equilibrium condition is smaller than the number of parameters that determine the spin configuration. 
This classical degeneracy can be lifted by the quantum fluctuation, which is most remarkable for the spin-1/2 case, and a specific spin state is selected as the ground state. The degeneracy lifting mechanism is called quantum order-by-disorder. 
Because the energy of the quantum fluctuation depends on the magnetic field, quantum phase transitions take place with varying magnetic field. 
A symbolic quantum effect is that the {\it up-up-down} state is stabilized in a finite field range, which results in a magnetization plateau at one-third of the saturation magnetization $M_{\rm s}$~\cite{Chubukov,Nikuni,Honecker,Farnell,Sakai,Hotta,Yamamoto1,Sellmann,Starykh}. This 1/3-magnetization plateau was clearly observed in Heisenberg-like TLAF Ba$_3$CoSb$_2$O$_9$~\cite{Shirata,Zhou,Susuki,Quirion,Koutroulakis}, and the entire quantum phases observed in magnetic fields were quantitatively explained using a microscopic model~\cite{Koutroulakis,Yamamoto2}. 

The honeycomb-lattice antiferromagnet (HLAF) with the nearest ($J_1$) and next-nearest neighbor ($J_2$) exchange interactions is a typical bond-frustrated magnet, in which the frustration arises from the competition between $J_1$ and $J_2$ interactions~\cite{Takano}. The ground state of the spin-1/2 $J_1\,{-}\,J_2$ Heisenberg HLAF at zero magnetic field has been attracting theoretical attention~\cite{Mosadeq,Ganesh,Bishop,Ganesh2}, mostly owing to the experiment on Bi$_3$Mn$_4$O$_{12}$(NO$_3$)~\cite{Matsuda}. Unconventional ground states including the spin liquid state are predicted. However, little is known about the ground state in a magnetic field even in the Ising-like case. 

\begin{figure*}[t]
\includegraphics[width=16 cm, clip]{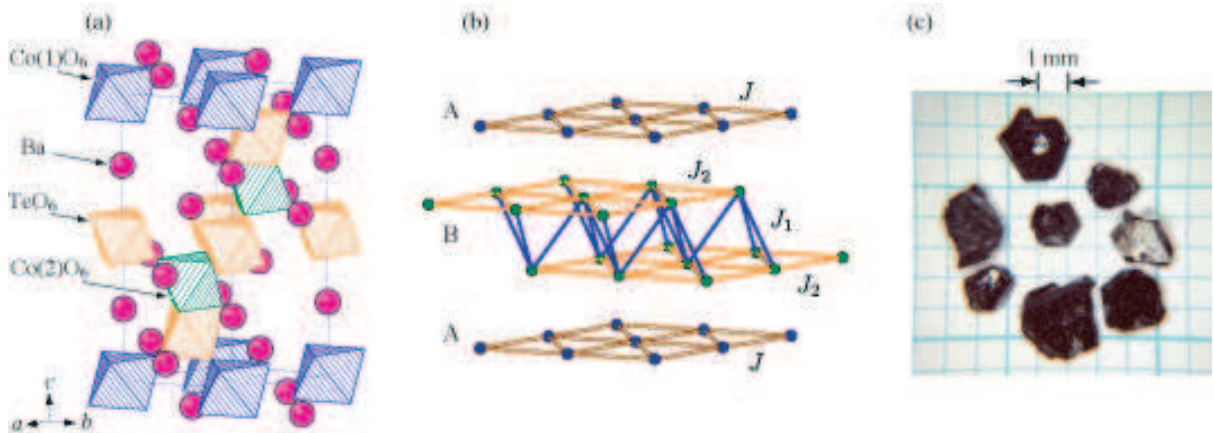}
\caption{(Color online) (a) Crystal structure of Ba$_2$CoTeO$_6$. The blue, green and orange octahedra are Co(1)O$_6$, Co(2)O$_6$, and TeO$_6$ octahedra, respectively. Dotted lines denote the chemical unit cell. (b) Magnetic subsystems A and B. Subsystem A is a uniform triangular lattice formed by Co(1) atoms. Subsystem B is composed of two uniform triangular lattices of Co(2) atoms, which are stacked with their lattice points mutually shifted to the other centers of triangles when projected onto the $ab$ plane. (c) Photograph of Ba$_2$CoTeO$_6$ single crystals. The wide plane is the crystallographic $c$ plane.
}
 \label{fig:structure}
\end{figure*}

Ba$_2$CoTeO$_6$ is a unique antiferromagnet that exhibits strong frustration that originates from both geometry and competing interactions. Ba$_2$CoTeO$_6$ crystallizes in a trigonal structure, $P{\bar 3}m$, as shown in Fig.~\ref{fig:structure}(a)~\cite{Ivanov}. There are two divalent cobalt sites, Co$^{2+}$(1) and Co$^{2+}$(2), with different octahedral environments. Co$^{2+}$(1) ions with effective spin-1/2 form a triangular lattice parallel to the $c$ plane, as shown in Fig.~\ref{fig:structure}(b), which we call subsystem A. Because a Co(1)O$_6$ octahedron is almost cubic, as observed in Ba$_3$CoSb$_2$O$_9$~\cite{Doi}, subsystem A is expected to be described as a spin-1/2 Heisenberg-like TLAF.
Co$^{2+}$(2) ions form a bilayer triangular lattice, as shown in Fig.~\ref{fig:structure}(b), which we call subsystem B. The lattice point of one triangular lattice shifts onto the center of the triangle of the other triangular lattice, when viewed along the $c$ axis. Because dominant superexchange interactions are considered to arise via TeO$_6$ octahedra linked with Co(2)O$_6$ octahedra by sharing corners, as discussed in Ref.~\cite{Yokota}, the interlayer exchange interaction $J_1$ and the nearest neighbor exchange interaction $J_2$ in the triangular lattice should be dominant. 

\begin{figure*}[htb]
\includegraphics[width=16 cm, clip]{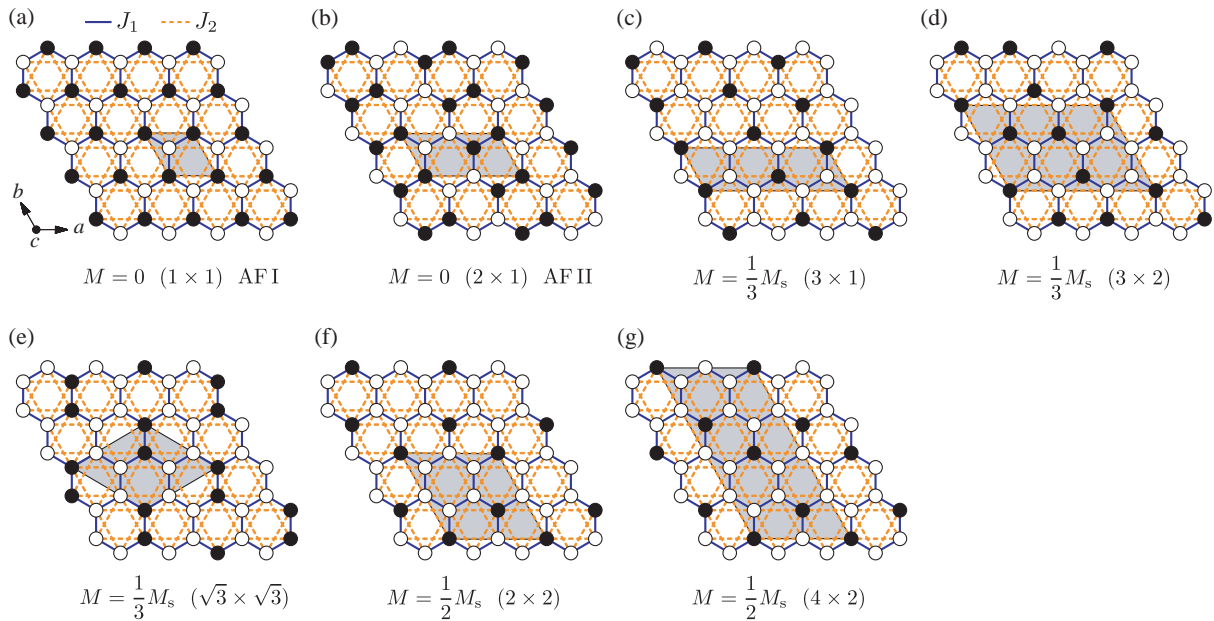}
\caption{(Color online) Spin structures at magnetization plateau states for subsystem B described as a $J_1\,{-}\,J_2$ Ising-like HLAF. 
Open and closed circles denote up and down spins, respectively. Shaded parallelograms are magnetic unit cells. (a) is a simple antiferromagnetic ordering on a hexagon (AF\,I). (b) is the $2\,{\times}\,1$ structure (AF\,II) observed at zero magnetic field~\cite{Ivanov}. (c), (d), and (e) are candidate structures for the $1/3$\,-\,plateau state, whereas (f) and (g) are those for the $1/2$\,-\,plateau state.}
 \label{fig:spin_structure}
\end{figure*}

A Co(2)O$_6$ octahedron is non-centrosymmetric. The sizes of two triangular faces perpendicular to the $c$ axis are different. The triangular face shared with a TeO$_6$ octahedron is smaller than the opposite face. 
Co$^{2+}$(2) shifts opposite to the TeO$_6$ octahedron. Consequently, the trigonal crystalline field acting on Co$^{2+}$(2) should be comparable to the spin-orbit coupling. Thus, it is considered that the exchange interaction between effective spins of Co$^{2+}$(2) ions is expressed by a strongly anisotropic XXZ model~\cite{Abragam,Lines}. The bilayer triangular lattice is equivalent to a honeycomb lattice, when projected onto the $c$ plane. Thus, subsystem B can be described as a spin-1/2 $J_1\,{-}\,J_2$ XXZ HLAF, as illustrated in Fig.~\ref{fig:spin_structure}.

It was reported that Ba$_2$CoTeO$_6$ undergoes antiferromagnetic ordering at around $T_{\rm N}\,{\simeq}\,15$ K~\cite{Ivanov,Mathieu}, and that spins are ordered parallel to the $c$ axis below $T_{\rm N}$. This indicates that the anisotropy in subsystem B is Ising-like. Figure \ref{fig:spin_structure}(a) shows the reported spin structure on subsystem B, in which the magnetic unit cell is enlarged to $2a\,{\times}\,a$ in the $c$ plane~\cite{Ivanov}. 

In this work, we performed magnetization and specific heat measurements using single crystals to investigate the ground-state properties and phase diagram in Ba$_2$CoTeO$_6$. 
It was found that successive phase transitions take place at $T_{\rm N1}\,{=}\,12.0$ and $T_{\rm N2}\,{=}\,3.0$ K, which correspond to the spin orderings on subsystems B and A, respectively. 
As shown below, subsystems A and B are approximately decoupled, so that the magnetization in Ba$_2$CoTeO$_6$ is given by the superposition of those for both subsystems. Therefore, we can observe the ground states and phase diagrams of the spin-1/2 Heisenberg-like TLAF and $J_1\,{-}\,J_2$ Ising-like HLAF separately in Ba$_2$CoTeO$_6$.

\section{Experimental Details}
Ba$_2$CoTeO$_6$ powder was first prepared via a chemical reaction 2BaCO$_3$\,{+}\,CoO\,{+}\,TeO$_2$\,{+}\,O$_2$ $\longrightarrow$ Ba$_2$CoTeO$_6$ + 2CO$_2$. Reagent-grade materials were mixed in stoichiometric quantities, and calcined at 1000 $^{\circ}$C for 24\,h in air. 
Ba$_2$CoTeO$_6$ single crystals were grown by the flux method. Ba$_2$CoTeO$_6$ powder and BaCl$_2$ were mixed in a molar ratio of 1\,{:}\,8 and placed into an alumina crucible. 
The crucible was covered with an alumina lid and placed in a box furnace. 
The temperature of the furnace was lowered from 1200 to 840 $^{\circ}$C over 240 h. Plate-shaped single crystals with a typical size of $2\,{\times}\,2{\times}\,0.3$ mm$^3$ were obtained, as shown in Fig.~\ref{fig:structure}(c). The wide plane of the crystals is the crystallographic $c$ plane. 

The magnetic susceptibilities of Ba$_2$CoTeO$_6$ single crystals were measured in the temperature range of $1.8\,{-}\,300$ K using a SQUID magnetometer (Quantum Design MPMS XL). 
The magnetization in a magnetic field of up to 60 T was measured at 4.2 and 1.3 K using an induction method with a multilayer pulse magnet at the Institute for Solid State Physics, The University of Tokyo. The absolute value of the high-field magnetization was calibrated with the magnetization measured using the SQUID magnetometer. The specific heat of Ba$_2$CoTeO$_6$ single crystals was measured down to 1.8 K in magnetic fields of up to 9 T using a physical property measurement system (Quantum Design PPMS) 
by the relaxation method. 

\begin{figure*}[t]
\includegraphics[width=14.0 cm, clip]{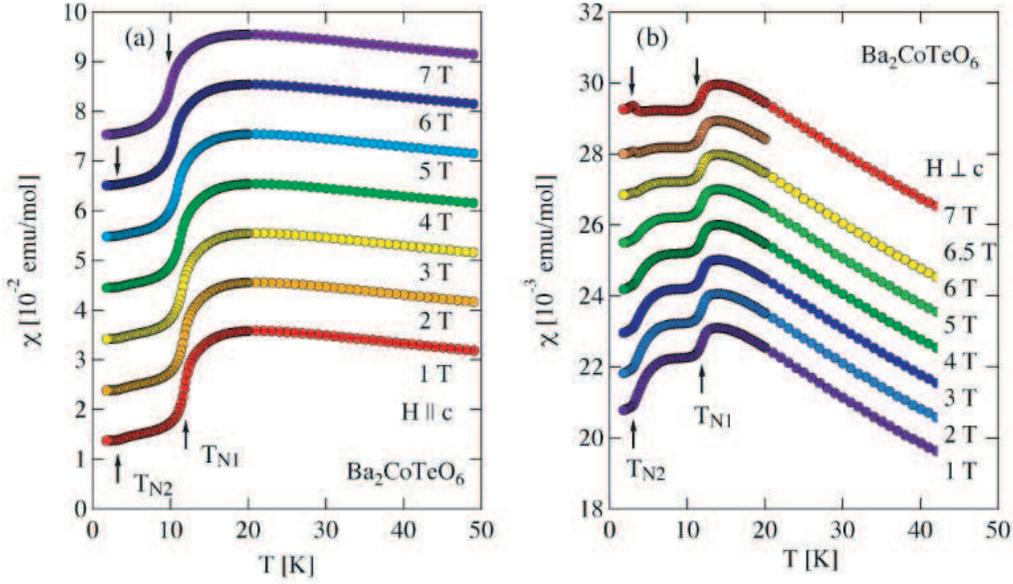}
\caption{(Color online) Magnetic susceptibilities (${\chi}\,{=}\,M/H$) in Ba$_2$CoTeO$_6$ measured at various magnetic fields (a) for $H\,{\parallel}\,c$ and (b) for $H\,{\perp}\,c$. Vertical arrows indicate magnetic phase transitions $T_{\rm N1}$ and $T_{\rm N2}$.
}
 \label{fig:chi}
\end{figure*}
\section{Results and Discussion}

Figure~\ref{fig:chi} shows the temperature dependences of magnetic susceptibilities (${\chi}\,{=}\,M/H$) measured at various magnetic fields (a) for $H\,{\parallel}\,c$ and (b) for $H\,{\perp}\,c$. Because the effective spin-1/2 description of the Co$^{2+}$ spin in an octahedral environment is valid only below liquid nitrogen temperature~\cite{Shirata}, we show the magnetic susceptibilities below 60 K. For $H\,{\parallel}\,c$, the magnetic susceptibility measured at $H\,{=}\,1$ T exhibits a rounded maximum at 20 K and an inflection point at $T_{\rm N1}\,{=}\,12.0$ K owing to magnetic ordering. With decreasing temperature, the magnetic susceptibility exhibits a bend anomaly at $T_{\rm N2}\,{=}\,3.0$ K indicative of the second magnetic ordering. The magnetic susceptibility measured at $H\,{=}\,1$ T for $H\,{\perp}\,c$ also shows the inflectional and bend anomalies at $T_{\rm N1}$ and $T_{\rm N2}$, respectively. With increasing magnetic field, $T_{\rm N1}$ for $H\,{\parallel}\,c$ shifts toward the low-temperature side, which is more clearly observed in specific heat data shown below. For $H\,{\perp}\,c$, the bend anomaly at $T_{\rm N2}$ observed below 5 T changes into a cusp anomaly above 6 T.

\begin{figure*}[t]
\includegraphics[width=14 cm, clip]{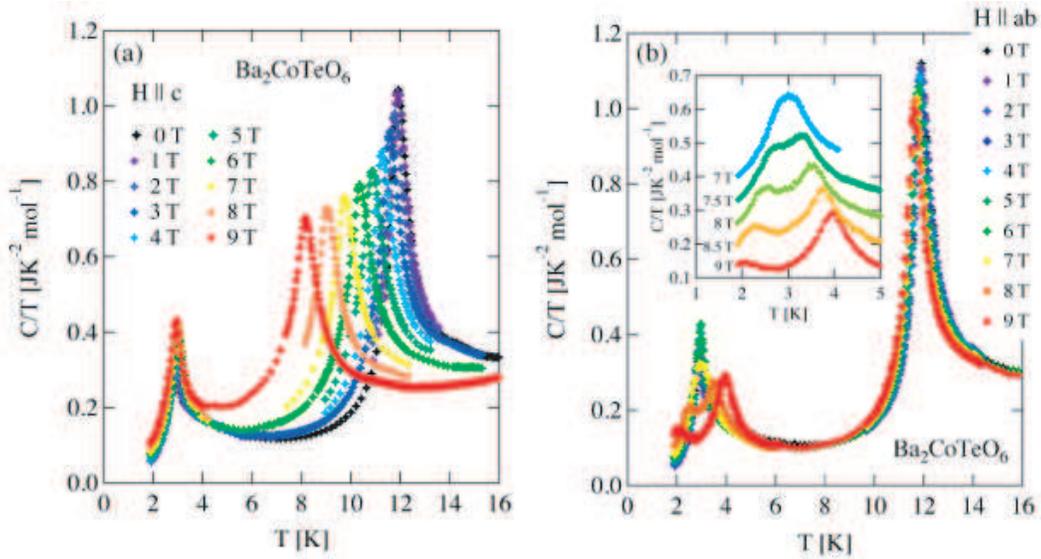}
\caption{(Color online) Specific heat divided by temperature of Ba$_2$CoTeO$_6$ at various magnetic fields (a) for $H\,{\parallel}\,c$ and (b) for $H\,{\perp}\,c$. The inset of (b) is the enlargement of specific heat between 1 and 5 K above $H\,{=}\,7$ T, where the data are shifted upward by multiples of 0.08 J/mol${\cdot}$K$^2$ with decreasing magnetic field.
}
 \label{fig:heat}
\end{figure*}

\begin{figure*}[ht]
\includegraphics[width=14 cm, clip]{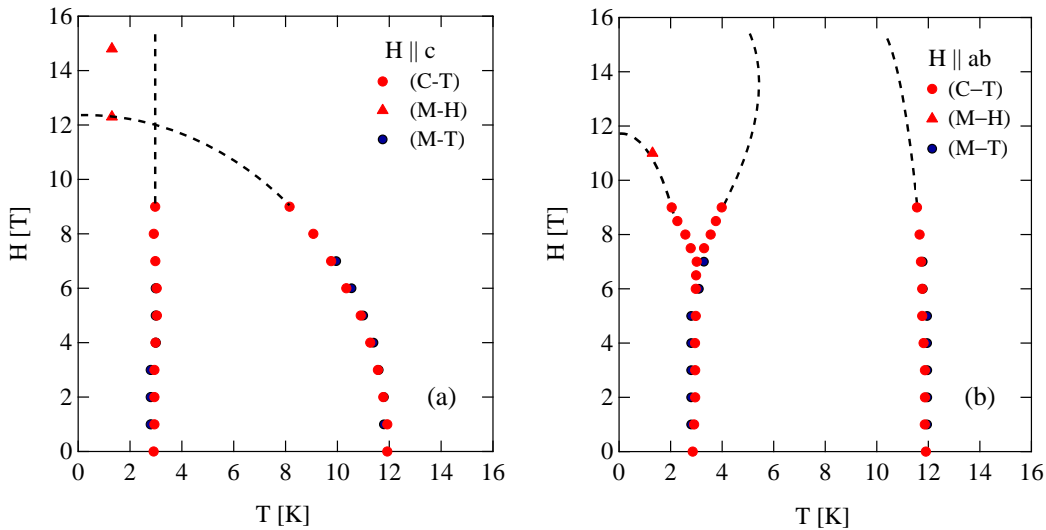}
\caption{(Color online) Magnetic field vs temperature phase diagrams in Ba$_2$CoTeO$_6$ (a) for $H\,{\parallel}\,c$ and (b) for $H\,{\perp}\,c$. Dashed lines are extrapolation of the phase boundaries. 
}
 \label{fig:phase}
\end{figure*}

Figure~\ref{fig:heat} shows the temperature dependence of the specific heat divided by temperature, $C/T$, below 16 K measured at various magnetic fields for $H\,{\parallel}\,c$ and $H\,{\perp}\,c$. At zero magnetic field, two sharp peaks indicative of magnetic phase transitions are observed at $T_{\rm N1}\,{=}\,11.93$ and $T_{\rm N2}\,{=}\,2.91$ K. $T_{\rm N1}$ is somewhat lower than $T_{\rm N}\,{\simeq}\,15$ K reported by Ivanov {\it et al.}~\cite{Ivanov}. For $H\,{\parallel}\,c$, $T_{\rm N1}$ shifts toward the low-temperature side with increasing magnetic field, whereas $T_{\rm N2}$ is almost independent of the magnetic field. For $H\,{\perp}\,c$, $T_{\rm N2}$ starts to split into two transitions at approximately 7 T with increasing magnetic field, whereas $T_{\rm N1}$ shifts slightly toward the low-temperature side. Figure~\ref{fig:phase} shows a summary of the transition data for both field directions. In Fig.~\ref{fig:phase}, transition data above 10 T were obtained from the high-field magnetization measurements shown below. The behavior of the phase boundaries related to $T_{\rm N2}$ is very similar to that observed in Ba$_3$CoSb$_2$O$_9$~\cite{Zhou,Quirion}.

\begin{figure}[t]
\includegraphics[width=8.0 cm, clip]{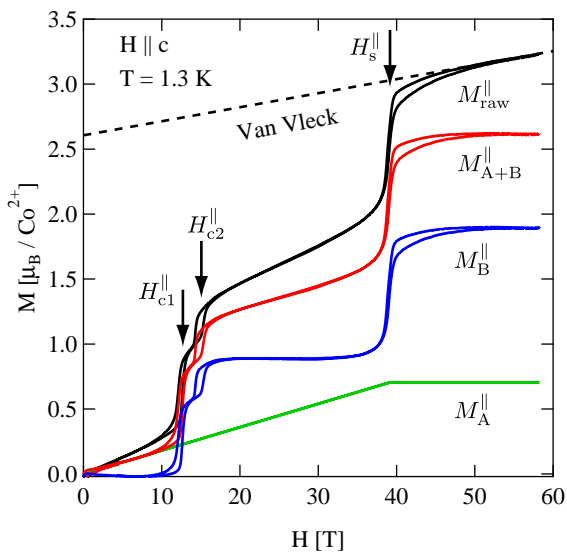}
\caption{(Color online) Magnetization process in Ba$_2$CoTeO$_6$ measured at 1.3\,K for $H\,{\parallel}\,c$. $M_{\rm raw}^{\parallel}$ is the raw magnetization. $M_{\rm A+B}^{\parallel}$ is the magnetization corrected for the Van Vleck paramagnetism, which is divided into two components $M_{\rm A}^{\parallel}$ and $M_{\rm B}^{\parallel}$ produced by spins in subsystems A and B, respectively. Arrows indicate the transition fields.}
 \label{fig:MH_para}
\end{figure}

\begin{figure}[h]
\includegraphics[width=8.0 cm, clip]{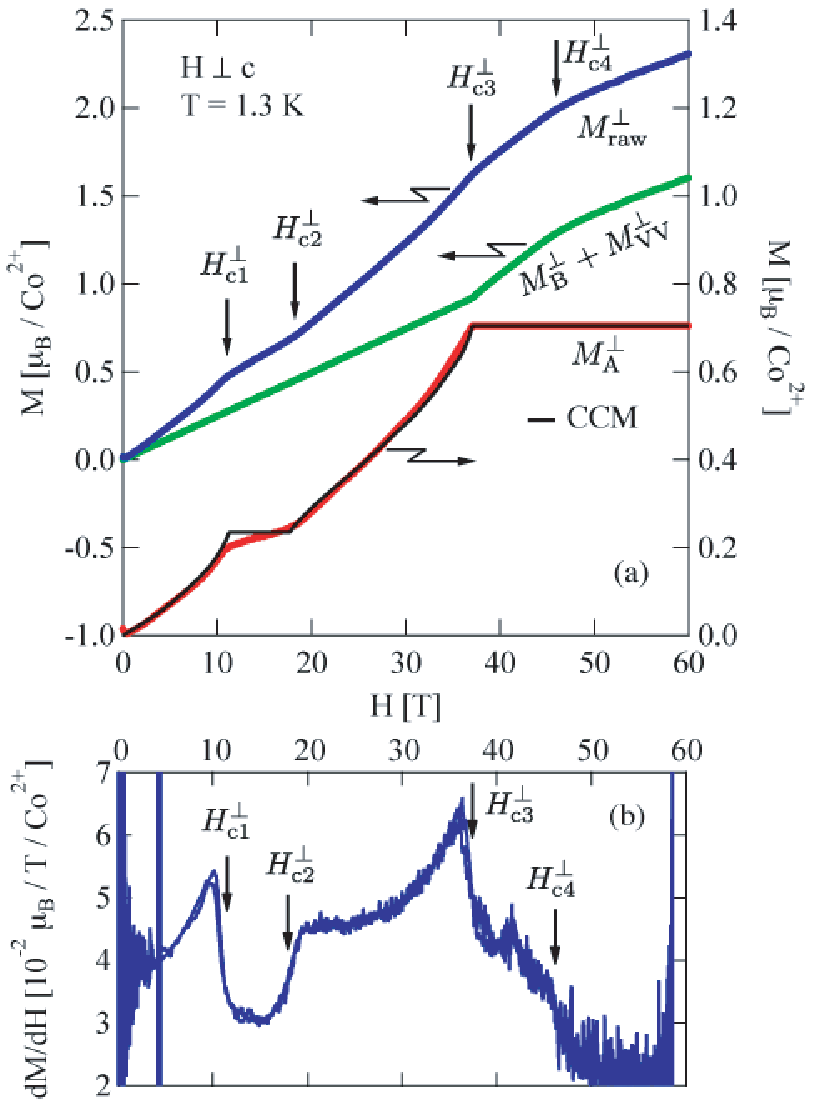}
\caption{(Color online) (a) Magnetization process in Ba$_2$CoTeO$_6$ measured at 1.3\,K for $H\,{\perp}\,c$. $M_{\rm raw}^{\perp}$ is the raw magnetization. $M_{\rm B}^{\perp}+M_{\rm VV}^{\perp}$ is the sum of the magnetizations of subsystem B and the Van Vleck paramagnetism. $M_{\rm A}^{\perp}$ is the magnetization of subsystem A. Vertical arrows indicate the transition fields. The solid line is the theoretical magnetization curve calculated by the higher order CCM~\cite{Farnell}. (b) $dM_{\rm raw}^{\perp}/dH$ measured at 1.3 K up to 58 T for $H\,{\perp}\,c$.}
 \label{fig:MH_perp}
\end{figure}

Figure~\ref{fig:MH_para} shows the magnetization process measured at 1.3 K for $H\,{\parallel}\,c$. Three transitions with a magnetization jump occur at $H_{\rm c1}^{\parallel}\,{=}\,12.3$ T, $H_{\rm c2}^{\parallel}\,{=}\,14.8$ T, and $H_{\rm s}^{\parallel}\,{=}\,39.0$ T. A small hysteresis is observed around these transitions. In the raw magnetization $M_{\rm raw}^{\parallel}$, the slopes for $H\,{<}\,H_{\rm c1}$ and $H_{\rm c2}\,{<}\,H\,{<}\,H_{\rm s}$ are almost the same, whereas, the magnetization slope for $H\,{>}\,H_{\rm s}$ is smaller than those for $H\,{<}\,H_{\rm s}$. This indicates that the magnetization produced by the effective spin-1/2 saturates at $H_{\rm s}$, and that the magnetization slope for $H\,{>}\,H_{\rm s}$ is attributed to the large temperature-independent Van Vleck paramagnetism characteristic of Co$^{2+}$ in octahedral environment. The  Van Vleck paramagnetic susceptibility for $H\,{\parallel}\,c$ is evaluated as ${\chi}_{\rm VV}^{\parallel}\,{=}\,6.09\,{\times}\,10^{-3}$ emu/mol. $M_{\rm A+B}^{\parallel}$ in Fig.~\ref{fig:MH_para} is the magnetization corrected for the Van Vleck paramagnetism. The saturation magnetization is obtained to be $M_{\rm s}^{\parallel}\,{=}\,2.60$ ${\mu}_{\rm B}$/Co$^{2+}$. 

For $M_{\rm A+B}^{\parallel}$, the magnetization slopes for $H\,{<}\,H_{\rm c1}^{\parallel}$ and $H_{\rm c2}^{\parallel}\,{<}\,H\,{<}\,H_{\rm s}^{\parallel}$ are almost the same. This suggests that $M_{\rm A+B}^{\parallel}$ is approximately given by the superposition of two components $M_{\rm A}^{\parallel}$ and $M_{\rm B}^{\parallel}$, where $M_{\rm A}^{\parallel}$ increases almost linearly in $H$ and saturates near 39 T, which is roughly similar to the magnetization curve for $H\,{\parallel}\,c$ in Ba$_3$CoSb$_2$O$_9$~\cite{Susuki}, and $M_{\rm B}^{\parallel}$ exhibits a stepwise magnetization process with plateaus at zero, one-third and one-half of the saturation magnetization $M_{\rm Bs}^{\parallel}$. It is natural to assume that $M_{\rm A}^{\parallel}$ and $M_{\rm B}^{\parallel}$ are the magnetizations of subsystems A and B, respectively, because the spins in subsystems A and B are expected to be Heisenberg-like and Ising-like, respectively. 
The $g$-factors for $H\,{\parallel}\,c$ in subsystems A and B are evaluated as $g_{\rm A}^{\parallel}\,{\simeq}\,4.22$ and $g_{\rm B}^{\parallel}\,{\simeq}\,5.66$, respectively.

Figure~\ref{fig:MH_perp} shows the magnetization process for $H\,{\perp}\,c$ measured at 1.3 K. In the raw magnetization data $M_{\rm raw}^{\perp}$ and $dM_{\rm raw}^{\perp}/dH$, four transitions are clearly observed at $H_{\rm c1}^{\perp}\,{=}\,11.0$ T, $H_{\rm c2}^{\perp}\,{=}\,18.0$ T, $H_{\rm c3}^{\perp}\,{=}\,37.2$ T and $H_{\rm c4}^{\perp}\,{=}\,45.9$ T. The magnetization obtained by extrapolating the magnetization slope above $H_{\rm c4}^{\perp}$ to zero magnetic field is approximately 1.0 ${\mu}_{\rm B}$/Co$^{2+}$, which is one-half of $M_{\rm s}^{\perp}\,{\simeq}\,2.0$ ${\mu}_{\rm B}$/Co$^{2+}$ expected as the saturation magnetization for $H\,{\perp}\,c$. Thus, $H_{\rm c4}^{\perp}$ is not the saturation field. 

As shown in Fig.~\ref{fig:MH_perp}(b), $dM_{\rm raw}^{\perp}/dH$ for $H\,{\leq}\,H_{\rm c3}^{\perp}$ is very similar to that observed for $H\,{\perp}\,c$ in the spin-1/2 Heisenberg-like TLAF Ba$_3$CoSb$_2$O$_9$ with small easy-plane anisotropy~\cite{Susuki}. Three critical fields $H_{\rm c1}^{\perp}$, $H_{\rm c3}^{\perp}$ and $H_{\rm c3}^{\perp}$ coincide with the lower and upper edge fields of the 1/3\,-\,magnetization plateau and the saturation field in Ba$_3$CoSb$_2$O$_9$ for $H\,{\perp}\,c$~\cite{Susuki} when we rescale the magnetic field. This indicates that subsystem A is approximately decoupled from subsystem B.

Assuming that the magnetization for the Ising-like subsystem B is linear in $H$ up to $H_{\rm c3}^{\perp}$, which is typical of the case for $H$ parallel to the hard-axis in three-dimensional Ising antiferromagnet~\cite{Kobayashi}, and the $g$-factor of subsystem A for $H\,{\perp}\,c$ is almost the same as $g_{\rm A}^{\parallel}\,{\simeq}\,4.22$, we divide $M_{\rm raw}^{\perp}$ into the magnetization of subsystem A ($M_{\rm A}^{\perp}$) and the sum of the magnetization of subsystem B and Van Vleck paramagnetic magnetization ($M_{\rm B}^{\perp}\,{+}\,M_{\rm VV}^{\perp}$), as shown in Fig.~\ref{fig:MH_perp}(a). $M_{\rm A}^{\perp}$ exhibits a 1/3\,-\,plateau caused by the quantum order-by-disorder~\cite{Chubukov,Nikuni,Honecker,Farnell,Sakai,Hotta,Yamamoto1,Sellmann,Starykh}. The solid line in Fig.~\ref{fig:MH_perp}(a) is the theoretical magnetization curve calculated by the higher order coupled cluster method (CCM)~\cite{Farnell} with $J/k_{\rm B}\,{=}\,23.5$ K and $g_{\rm A}^{\perp}\,{=}\,4.22$. The magnetization $M_{\rm A}^{\perp}$ is in good quantitative agreement with the theoretical result. From these results, we infer that the spins in subsystem A are ordered parallel to the $ab$ plane at $T_{\rm N2}$ and paramagnetic above $T_{\rm N2}$, and that the spins in subsystem B are ordered parallel to the $c$ axis at $T_{\rm N1}$, although Ivanov {\it et al.}~\cite{Ivanov} reported that all the spins are ordered at $T_{\rm N}\,{\simeq}\,15$ K along the $c$ axis. 

The results of high-field magnetization measurements show that the total magnetization is approximately given by the superposition of magnetizations for isolated subsystems A and B. This indicates that the coupling between the two subsystems is weak. It is considered that the anomalies at $H_{\rm c3}^{\perp}\,{=}\,37.2$ and $H_{\rm c4}^{\perp}\,{=}\,45.9$ T in $M_{\rm B}^{\perp}\,{+}\,M_{\rm VV}^{\perp}$ for $H\,{\perp}\,c$ are attributed to the phase transitions in subsystem B. Usually, the magnetization curve for the classical Ising-like magnet is linear in $H$ and displays no transition up to the saturation when the magnetic field is applied parallel to the hard axis. Therefore, we speculate that these transitions are the quantum phase transitions due to the transverse magnetic field in the $J_1\,{-}\,J_2$ Ising-like HLAF. The transitions at $H_{\rm c3}^{\perp}\,{=}\,37.2$ T for $H\,{\perp}\,c$ and at $H_{\rm s}^{\parallel}\,{=}\,39.0$ T for $H\,{\parallel}\,c$ occur simultaneously in both subsystems. If the interaction between the subsystems is negligible, then these transitions take place independently. It is considered that the transitions that take place originally at slightly different magnetic fields in these two subsystems occur simultaneously with the help of the weak exchange interaction between the subsystems. 

Next, we examine the ground state for $H\,{\parallel}\,c$ in subsystem B, assuming the $J_1\,{-}\,J_2$ Ising HLAF.  If $J_1$ is much larger than $J_2$, a simple antiferromagnetic ordering on a hexagon (AF\,I) takes place, as shown in Fig.~\ref{fig:spin_structure}(a). However, the spin state observed below $T_{\rm N1}$ is as shown in Fig.~\ref{fig:spin_structure}(b) with a unit cell enlarged to $2a\,{\times}\,a$ (AF\,II)~\cite{Ivanov}. Because the energies of AF\,I and AF\,II per spin are expressed as $E^{(\rm a)}=-(3/8)J_1+(3/4)J_2$ and $E^{(\rm b)}=-(J_1+2J_2)/8$, respectively, it is concluded that $J_1\,{<}\,4J_2$ in Ba$_2$CoTeO$_6$. 

\begin{figure}[t]
\includegraphics[width=8.0 cm, clip]{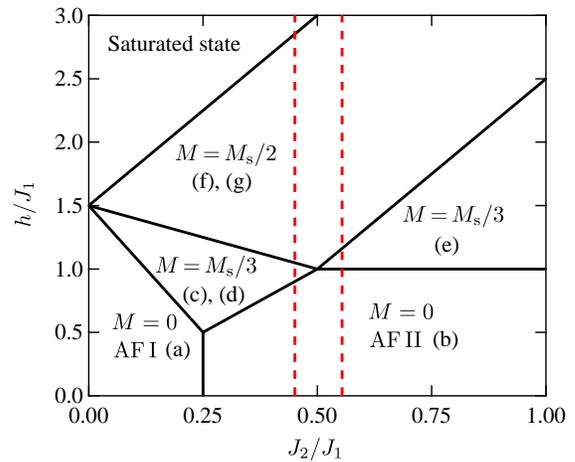}
\caption{(Color online) Ground-state phase diagram of the $J_1\,{-}\,J_2$ Ising HLAF model in magnetic fields. The dashed lines are the ground states for $J_2/J_1\,{=}\,0.45$ and 0.55, which were obtained for Ba$_2$CoTeO$_6$.}
\label{fig:phase_Ising}
\end{figure}

According to the Kanamori theory~\cite{Kanamori2}, the stable state just below the saturation field is such that the density of down spins is maximum under the condition that no two down spins interact via given exchange interactions. The spin states shown in Figs.~\ref{fig:spin_structure}(f) and (g) satisfy this condition and have the maximum magnetization of $M\,{=}\,M_{\rm Bs}^{\parallel}/2$. These two states have the same energy $E^{(\rm f,g)}=-h/4$ with $h\,{=}\,g{\mu}_{\rm B}H$. Because any sequences of (f) and (g) structures in the $b$ direction with the same pattern in the $a$ direction have the same energy, the spin state of the 1/2\,-\,plateau state is infinitely degenerate. Comparing $E^{(\rm f,g)}$ with the energy of the saturated state given by $E^{(\rm s)}=(3/8)(J_1+2J_2)-h/2$, the saturation field is obtained as $h_{\rm s}=(3/2)J_1+3J_2$.

The spin structures shown in Figs.~\ref{fig:spin_structure}(c)\,$-$\,(e) are candidates of the 1/3\,-\,plateau state. Structures in Figs.~\ref{fig:spin_structure}(c) and (d) have the same energy expressed as $E^{(\rm c,d)}=-(3J_1-2J_2)/24-h/6$, whereas the energy of structure in Fig.~\ref{fig:spin_structure}(e) is given by $E^{(\rm e)}=(J_1-6J_2)/24-h/6$. The structures Figs.~\ref{fig:spin_structure}(c) and (d) are stable for $J_1\,{\geq}\,2J_2$, whereas, the structure in Fig.~\ref{fig:spin_structure}(e) is stable for $J_1\,{<}\,2J_2$. Note that the 1/3\,-\,plateau state for $J_1\,{\geq}\,2J_2$ is infinitely degenerate, because any sequence of (c) and (d) structures in the $b$ direction have the same energy. The critical field $h_{\rm c2}$ values are obtained as $h_{\rm c2}=\left(3J_1-2J_2\right)/2$ and $\left(6J_2-J_1\right)/2$ for $J_1\,{\geq}\,2J_2$ and $J_1\,{<}\,2J_2$, respectively. Comparing $E^{(\rm c,d)}$ and $E^{(\rm e)}$ with the energy of the zero-field ground state $E^{(\rm b)}=-(J_1+2J_2)/8$, the critical field $h_{\rm c1}$ is obtained as $h_{\rm c1}=2J_2$ and $J_1$ for $J_1\,{\geq}\,2J_2$ and $J_1\,{<}\,2J_2$, respectively. Figure~\ref{fig:phase_Ising} shows the ground state phase diagram in the $J_2/J_1\,{-}\,h/J_1$ plane.

Using $H_{\rm c1}^{\parallel}\,{=}\,12.3$ T, $H_{\rm s}^{\parallel}\,{=}\,39.0$ T and $g_{\rm B}^{\parallel}\,{\simeq}\,5.66$, we obtain $J_1\,{\simeq}\,52$ K and $J_2\,{\simeq}\,24$ K for $J_1\,{\geq}\,2J_2$, and $J_1\,{\simeq}\,47$ K and $J_2\,{\simeq}\,26$ K for $J_1\,{<}\,2J_2$. 
The dashed lines in Fig.~\ref{fig:phase_Ising} are the ground states for these two sets of the parameter $J_2/J_1\,{=}\,0.45$ and 0.55. 
From the present experiments, we cannot determine which parameter is realized in Ba$_2$CoTeO$_6$. Using these parameters, the second critical field is calculated as $H_{\rm c2}^{\rm cal}\,{=}\,14.3$ T, which is consistent with $H_{\rm c2}^{\rm exp}\,{=}\,14.8$ T observed in this experiment. This confirms that subsystem B is described as the $J_1\,{-}\,J_2$ Ising HLAF is approximately isolated from subsystem A. The reason that the field range of the 1/3\,-\,plateau state is small is because $J_1$ is close to $2J_2$.

\section{Conclusion}
We have presented the results of specific heat and magnetization measurements of Ba$_2$CoTeO$_6$. It was found that Ba$_2$CoTeO$_6$ is composed of two approximately isolated subsystems A and B that are described as a spin\,-\,1/2 Heisenberg-like TLAF with small easy-plane anisotropy and $J_1\,{-}\,J_2$ Ising-like HLAF, respectively. Ba$_2$CoTeO$_6$ exhibits two phase transitions, $T_{\rm N1}\,{\simeq}\,12.0$\,K and $T_{\rm N2}\,{\simeq}\,3.0$\,K, which correspond to the orderings of subsystems B and A, respectively. For $H\,{\perp}\,c$, the magnetization process of subsystem A is in good quantitative agreement with the theoretical result for spin\,-\,1/2 Heisenberg TLAF~\cite{Farnell}. The stepwise magnetization process for subsystem B for $H\,{\parallel}\,c$ can be understood within the framework of $J_1\,{-}\,J_2$ Ising HLAF. However, the spin states of the 1/2\,- and 1/3\,-\,plateaus for $J_1\,{\geq}\,2J_2$ are infinitely degenerate. These degeneracies can be lifted by the quantum fluctuation that originates from the finite transverse component of the exchange interactions. It is interesting to investigate how these plateau states change with increasing the magnitude of the transverse component.

\begin{acknowledgments}
We express our sincere thanks to J. Richter for fruitful discussions and comments. This work was supported by Grants-in-Aid for Scientific Research (A) No. 26247058 and Young Scientists (B) No. 26800181 from the Japan Society for the Promotion of Science.
\end{acknowledgments}

\end{document}